\documentclass[iop]{emulateapj}
\usepackage[flushleft]{threeparttable}
\usepackage[pdftex,usenames,dvipsnames]{xcolor}
\usepackage{natbib}
\usepackage{url}
\usepackage[none]{hyphenat}
\usepackage{times}
\usepackage[linktocpage=true,backref=true,pagebackref=false,bookmarks=true,bookmarksnumbered=False,colorlinks=true,linkcolor=NavyBlue,citecolor=NavyBlue,urlcolor=NavyBlue]{hyperref}
\usepackage[all]{hypcap}
\usepackage{mathtools}  
\pdfpageattr {/Group << /S /Transparency /I true /CS /DeviceRGB>>}  
\usepackage{booktabs}  
\usepackage{amsmath}   

\shorttitle{A population of IMBHs in dwarf galaxies up to z=1.5}
\shortauthors{Mezcua et al.}

\begin{document}

\title{A population of intermediate-mass black holes \\in dwarf starburst galaxies up to redshift=1.5}

\author{M.~Mezcua\altaffilmark{1,2}, F.~Civano\altaffilmark{3,1}, G. Fabbiano\altaffilmark{1}, T. Miyaji\altaffilmark{4,5} and S. Marchesi\altaffilmark{3,1}}


\affil{$^{1}$Harvard-Smithsonian Center for Astrophysics (CfA), 60 Garden Street, Cambridge, MA 02138, USA; marmezcua.astro@gmail.com}
\affil{$^{2}$D\'epartement de Physique, Universit\'e de Montr\'eal, C.P. 6128, Succ. Centre-Ville, Montr\'eal, QC H3C 3J7, Canada} 
\affil{$^{3}$Yale Center for Astronomy and Astrophysics, 260 Whitney ave., New Haven, CT 06520-8121, USA} 
\affil{$^{4}$Instituto de Astronom\'ia sede Ensenada, Universidad Nacional Aut\'onoma de M\'exico, Km. 103, Carret. Tijuana-Ensenada, Ensenada, BC 22860, Mexico}
\affil{$^{5}$University of California San Diego, Center for Astrophysics and Space Sciences, 9500 Gilman Drive, La Jolla, CA 92093-0424, USA}

\begin{abstract}
We study a sample of $\sim$50,000 dwarf starburst and late-type galaxies drawn from the COSMOS survey with the aim of investigating the presence of nuclear accreting black holes (BHs) as those seed BHs from which supermassive BHs could grow in the early Universe. We divide the sample into five complete redshift bins up to $z=1.5$ and perform an X-ray stacking analysis using the \textit{Chandra} COSMOS-Legacy survey data. After removing the contribution from X-ray binaries and hot gas to the stacked X-ray emission, we still find an X-ray excess in the five redshift bins that can be explained by nuclear accreting BHs. This X-ray excess is more significant for $z<0.5$. At higher redshifts, these active galactic nuclei could suffer mild obscuration, as indicated by the analysis of their hardness ratios. 
The average nuclear X-ray luminosities in the soft band are in the range 10$^{39}-10^{40}$ erg s$^{-1}$. Assuming that the sources accrete at $\geq$ 1\% the Eddington rate, their BH masses would be $\leq$ 10$^{5}$ M$_{\odot}$, thus in the intermediate-mass BH regime, but their mass would be smaller than the one predicted by the BH-stellar mass relation. If instead the sources follow the correlation between BH mass and stellar mass, they would have sub-Eddington accreting rates of $\sim$ 10$^{-3}$ and BH masses 1-9 $\times$ 10$^{5}$ M$_{\odot}$. We thus conclude that a population of intermediate-mass BHs exists in dwarf starburst galaxies, at least up to $z$=1.5, though their detection beyond the local Universe is challenging due to their low luminosity and mild obscuration unless deep surveys are employed. 
\end{abstract}


\section{Introduction}
Two main scenarios have been proposed for the formation and accretion of supermassive black holes (SMBHs). SMBHs could have formed from the first generation (Population III) of $\sim$100 M$_{\odot}$ stellar seeds (e.g. \citealt{2010A&ARv..18..279V} and references therein). Alternatively, SMBHs in the early Universe could grow from heavier $M_\mathrm{BH} \sim 10^{4-6}$ M$_{\odot}$ seed BHs formed by direct collapse of pristine gas in primordial halos (e.g. \citealt{1995ApJ...443...11E}). Such halos should be close (within 15 kpc) to protogalaxies of $\sim 10^{7}$ M$_{\odot}$ emitting a high Lyman-Werner radiation in order to avoid gas fragmentation and the formation of stars (e.g. \citealt{2014MNRAS.443..648A}). 
Observationally, the finding of an increasing number of SMBHs at redshifts $z\sim$7 (e.g. \citealt{2001AJ....122.2833F,2003AJ....125.1649F}; \citealt{2007AJ....134.2435W,2010AJ....139..906W}; \citealt{2011Natur.474..616M}; \citealt{2013ApJ...779...24V}; \citealt{2015Natur.518..512W}) poses a challenge to the above scenarios. Population III remnant BHs would have to constantly accrete at nearly Eddington rates, undergo short phases of super-Eddington accretion (\citealt{2005ApJ...633..624V}), or be embedded in a nuclear star cluster and fed by flows of dense cold gas (\citealt{2014Sci...345.1330A}) in order to reach BH masses $\geq 10^{9}$ M$_{\odot}$ in less than 1 Gyr (see review by \citealt{2014GReGr..46.1702N}).

Proving the existence of the primordial BHs at z $>$ 10 is extremely challenging with the current instrumentation (e.g. \citealt{2015ApJ...808..139S}); however, a discrimination between the different BH formation scenarios can be performed by studying the population of low-mass BHs in the local Universe. Such low-mass BHs, also known as intermediate-mass BHs (IMBHs), with masses in the range $10^{2}-10^{6}$ M$_{\odot}$, are expected to reside in low-mass, star-forming dwarf galaxies that have not significantly grown through mergers/accretion, resembling  the primordial galaxies formed in the early Universe. Given that 'light' Pop III seed BHs are expected to be more abundant than the 'massive' direct collapse BHs in the infant Universe (e.g. \citealt{2010MNRAS.408.1139V}), a large BH occupation fraction ($\sim$90\%) should be found in today's dwarf galaxies if Pop III remnants were the dominant seeding mechanism (e.g. \citealt{2010MNRAS.408.1139V}; \citealt{2010A&ARv..18..279V}; \citealt{2012NatCo...3E1304G}). Probing the fraction of BHs in low-mass galaxies, their mass density and host galaxy properties is thus pivotal to discern between the different models describing the formation of the first BHs (e.g. \citealt{2010A&ARv..18..279V}; \citealt{2012NatCo...3E1304G}). 

The first systematic searches for low-mass BHs focused on optically selected samples of active galactic nuclei (AGN) with $M_\mathrm{BH} < 10^{6}$ M$_{\odot}$, as derived from the width of optical emission lines (e.g.  \citealt{2004ApJ...607...90B,2008AJ....136.1179B}; \citealt{2004ApJ...610..722G,2007ApJ...670...92G}; \citealt{2005ApJ...632..799P}; \citealt{2013ApJ...775..116R}). More than 200 candidates were found, nearly a quarter of which have been confirmed as accreting BHs according to their X-ray emission ($L_\mathrm{0.5-2keV} = 10^{41}-10^{43}$ erg s$^{-1}$; \citealt{2009ApJ...698.1515D}; \citealt{2012ApJ...755..167D}). The detection of unresolved X-ray emission, in some cases spatially coincident with jet/core radio emission, constitutes the most compelling signature of an accreting BH in the absence of dynamical mass measurements (typically limited to the Local Group in the case of dwarf galaxies, e.g. \citealt{2005ApJ...628..137V}; \citealt{2009ApJ...699L.113L}; \citealt{2010ApJ...714..713S,2014Natur.513..398S}; \citealt{2015ApJ...809..101D}) and has provided further evidence for their presence in a few more tens of low-mass galaxies (e.g. \citealt{2008ApJ...688..794S}; \citealt{2011Natur.470...66R,2014ApJ...787L..30R}; \citealt{2012ApJ...750L..24R}; \citealt{2013ApJ...773..150S}; \citealt{2014ApJ...782...55Y}; \citealt{2015ApJ...809L..14B}; \citealt{2015ApJ...805...12L}; \citealt{2015ApJ...798...38S}; \citealt{2015ApJ...806...37W}). Additional searches for IMBHs have been performed in the infrared band (e.g. \citealt{2007ApJ...663L...9S,2008ApJ...677..926S,2014ApJ...784..113S}; \citealt{2014arXiv1411.3844M}; \citealt{2015MNRAS.454.3722S}), or using globular clusters (e.g. \citealt{2005ApJ...634.1093G}; \citealt{2010ApJ...710.1063V}; \citealt{2011A&A...533A..36L}) and also ultraluminous X-ray sources (ULXs; e.g. \citealt{2009Natur.460...73F}; \citealt{2011AN....332..379M}; \citealt{2012MNRAS.423.1154S}; \citealt{2013MNRAS.436.1546M}; \citealt{2014Natur.513...74P}). Among ULXs, the two best IMBH candidates have spatially coincident X-ray and radio emissions and are suggested to be the nucleus of a dwarf galaxy stripped in the course of minor merger (\citealt{2012ApJ...747L..13F}; \citealt{2013MNRAS.436.3128M,2015MNRAS.448.1893M}; \citealt{2013ApJ...768L..22S}). 

The presence of accreting BHs in low-mass early-type galaxies is expected from $M_\mathrm{BH}$ scaling relations and has already been inferred in several cases (e.g. \citealt{2010ApJ...714...25G}; \citealt{2012ApJ...747...57M,2015ApJ...799...98M}; \citealt{2014ApJ...790...16C}; \citealt{2015arXiv150703170P}). Although it may be easier to study BH accretion in redder galaxies where enhanced X-ray emission from high star-forming phenomena (e.g. \citealt{2012MNRAS.419.2095M,2012MNRAS.426.1870M}) is not significant, certainly more searches have been completed among late-type spirals, for which several individual cases have been reported (e.g. \citealt{2005ApJ...632..799P}; \citealt{2007ApJ...663L...9S}; \citealt{2009ApJ...700.1759G}; \citealt{2011Natur.470...66R}; \citealt{2015ApJ...809L..14B}; \citealt{2015ApJ...805...12L}). However, observational evidence for the presence of a population of IMBHs in dwarf starburst galaxies (irregular and spiral) is still scant. The largest IMBH samples are skewed toward broad-line AGN with high Eddington ratios ($L_\mathrm{bol}/L_\mathrm{Edd}>0.1$; see \citealt{2014ApJ...782...55Y}), are not complete, or cover small volumes ($z<0.3$). 
To circumvent these biases, we have performed a search for accreting BHs in low-mass starburst and late-type galaxies in the COSMOS survey (\citealt{2007ApJS..172....1S}), which is the only large (2 deg$^{2}$) survey for which a complete, deep ($i_\mathrm{AB}\sim$26.5), multiwavelength dataset exists and that all major telescopes have deeply observed (e.g. space: \textit{Chandra, Hubble, Spitzer, Herschel, GALEX, XMM-Newton, NuSTAR}; ground: VLA, Subaru, Canada-France-Hawaii Telescope, Magellan, VLT, VISTA). To identify signs of BH accretion we make use of the recently completed \textit{Chandra} COSMOS-Legacy survey (\citealt{2016arXiv160100941C}), the combination of the C-COSMOS survey (\citealt{2009ApJS..184..158E}; \citealt{2012ApJS..201...30C}) and a new X-ray Visionary Program project approved in \textit{Chandra} Cycle 14. \textit{Chandra} COSMOS-Legacy covers the whole 2.2 deg$^{2}$ of the COSMOS field with ACIS-I imaging at a depth of $\sim$150 ks. In this paper, we present the finding of a population of accreting BHs in low-mass starburst and spiral galaxies up to z = 1.5 based on the X-ray stacking analysis performed on \textit{Chandra} COSMOS-Legacy. The results of $\sim$200 actual X-ray detections up to z $\sim$ 4 will be reported in a future paper (M. Mezcua et al. in preparation). The paper is divided as follows: in Sections~\ref{sample} and \ref{analysis} we describe the data and analysis; the results obtained are reported and discussed in Section~\ref{results}. Final conclusions are given in Section~\ref{conclusions}. Throughout the paper we adopt a $\Lambda$CDM cosmology with parameters $H_{0}=70$ km s$^{-1}$ Mpc$^{-1}$, $\Omega_{\Lambda}=0.73$ and $\Omega_{m}=0.27$.

\begin{figure}
 \includegraphics[width=0.53\textwidth]{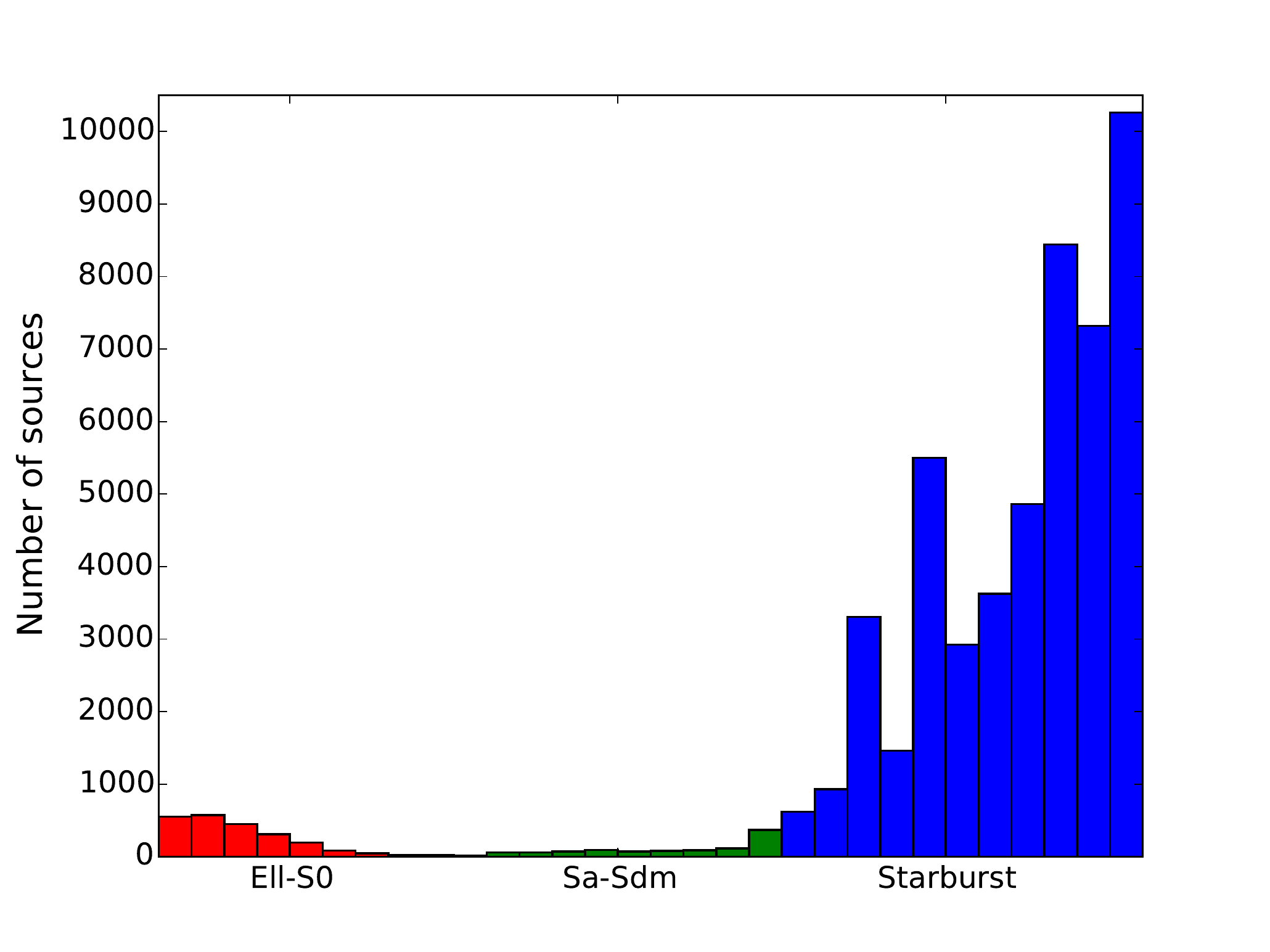}
  \includegraphics[width=0.49\textwidth]{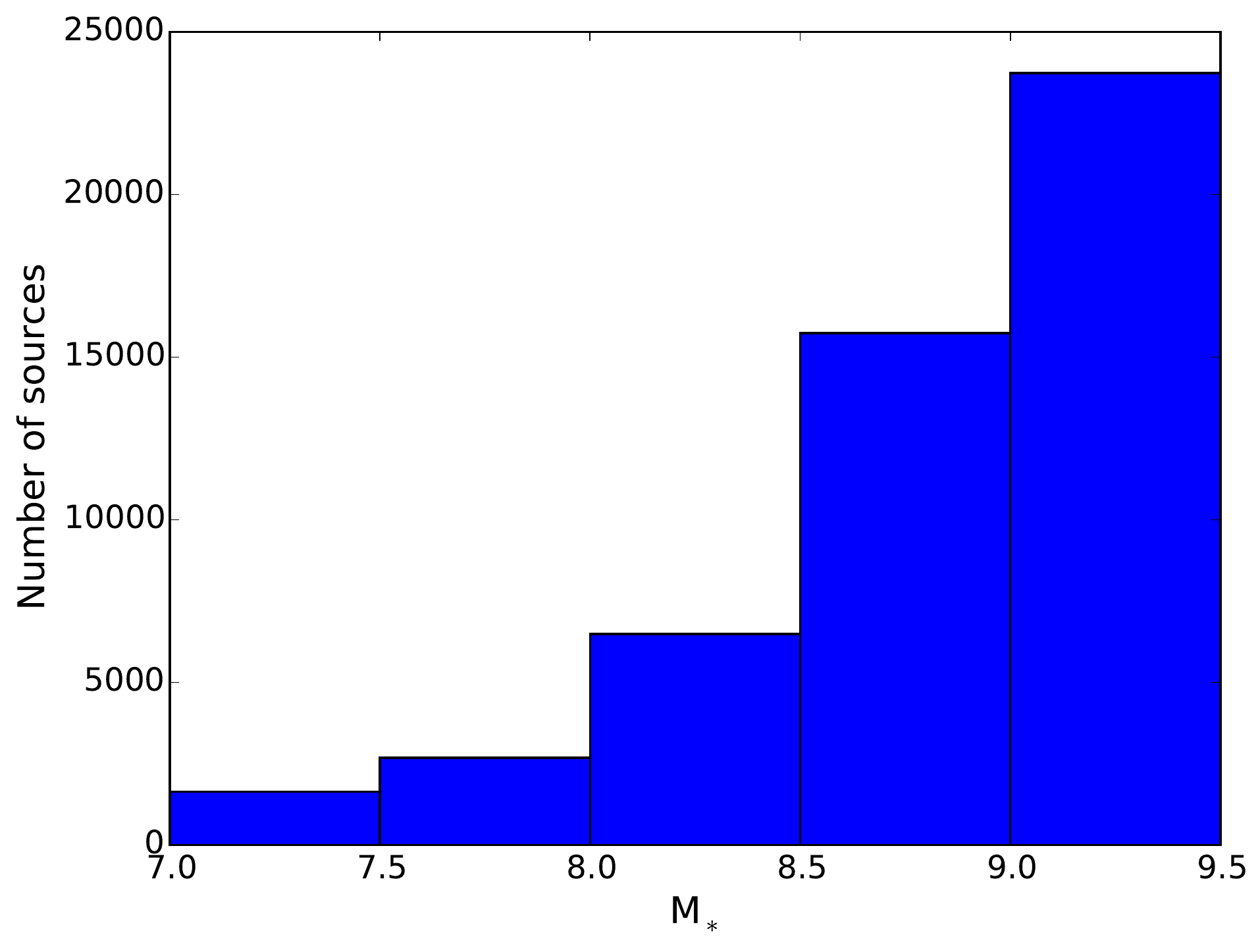}
\protect\caption[morphology]{\textbf{Top}: distribution of morphologies for the sample of low-mass galaxies. Early-type galaxies are shown in red, spiral galaxies in green, starburst in blue. \textbf{Bottom}: distribution of stellar masses for the sample of non-elliptical dwarf galaxies.}
\label{morphology}
\end{figure}

\section{Sample selection}
\label{sample}
The sample of low-mass star-forming galaxies is selected from a recent version of the COSMOS photometric catalog of \cite{2009ApJ...690.1236I,2010ApJ...709..644I} including the four Ultra-VISTA YJHK bands from \cite{2012A&A...544A.156M}. Photometric redshifts and galaxy properties (mass, age, star formation rate (SFR), and galaxy type) are derived from the fit of the spectral energy distribution (SED) using elliptical, spiral and star-forming galaxy templates from \cite{2007ApJ...663...81P} and \cite{2003MNRAS.344.1000B} and assuming a \citet{2003PASP..115..763C} initial mass function (see \citealt{2009ApJ...690.1236I,2010ApJ...709..644I} for details). The AGN contribution is negligible according to the SED fitting and is thus not taken into account in the estimates of the galaxy properties. 

\begin{figure}
 \includegraphics[width=0.48\textwidth]{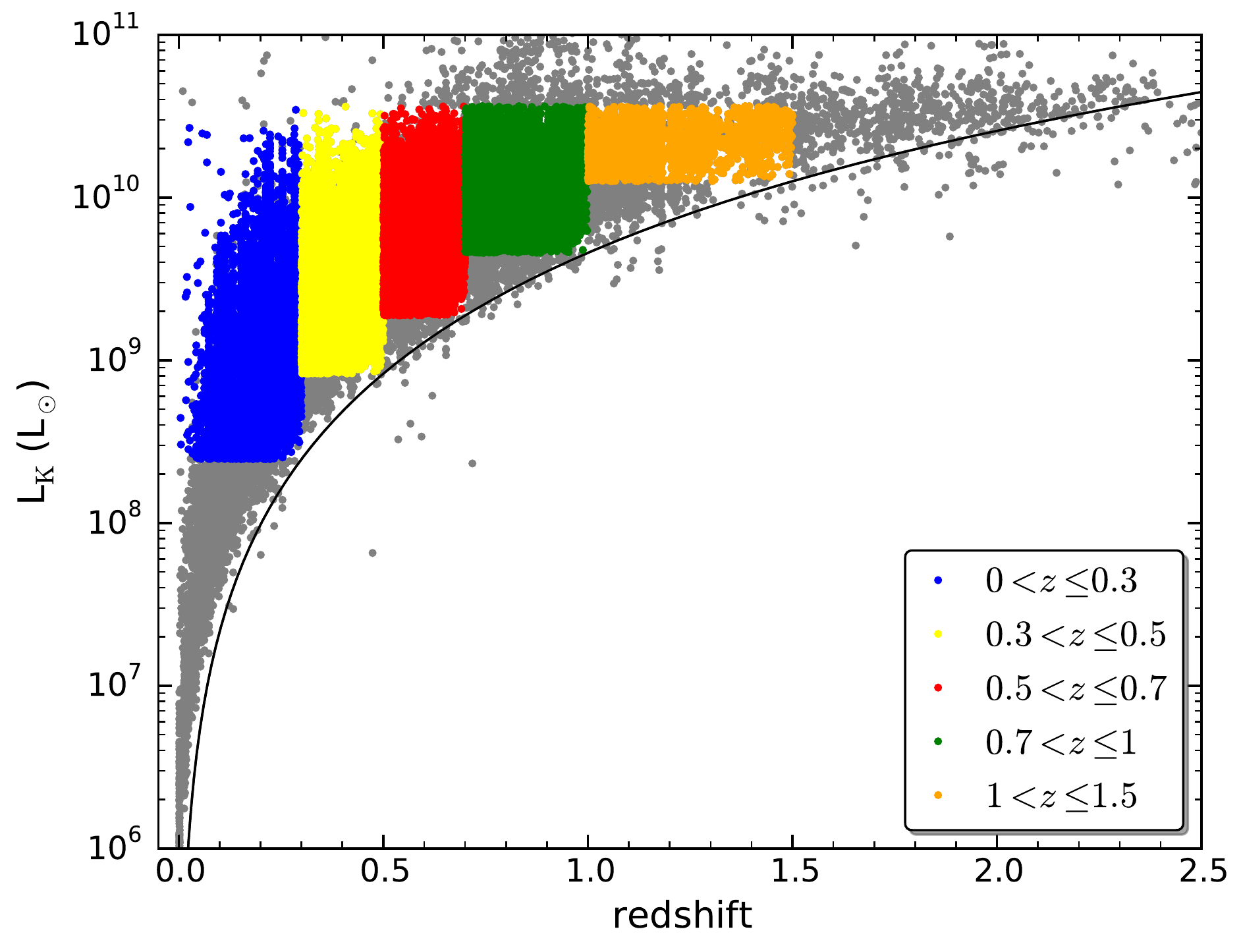}
\protect\caption[completeness]{\textit{K}-band luminosity vs. redshift for the selected sample of low-mass galaxies ($10^{7}\leq M_{*}\leq 3 \times 10^{9}$ M$_{\odot}$).  The solid black line represents the \textit{K}-band sensitivity limit of 23 mag. The color-coded regions show the complete redshift bins considered in the stacking analysis.}
\label{completeness}
\end{figure}

We select low-mass galaxies as being brighter than 24 mag in the \textit{i}-band, since the photometric redshift precision is $\Delta z/(1+z)=0.012$ at $i_\mathrm{AB}<24$ (\citealt{2009ApJ...690.1236I}), and as having a stellar mass $M_{*}\leq3 \times 10^{9}$ M$_{\odot}$ (or log $M_{*} \leq 9.5$ M$_{\odot}$, e.g. \citealt{2013ApJ...773..150S}). We exclude all those galaxies masked in the Ilbert et al. catalog as not having reliable photometric redshifts or stellar masses (i.e. $M_{*} < 10^{7}$ M$_{\odot}$). With these criteria, we find 52508 galaxies with $10^{7}\leq M_{*}\leq3 \times 10^{9}$ M$_{\odot}$. The SED fitting provides a classification type (Ell-S0, Sa-Sc, Sd-Sdm, starburst; see Figure~\ref{morphology}, top) for each low-mass galaxy. We find that most galaxies (93.8\%) are fitted by starburst templates according to the SED shape, with a smaller fraction (4.2\%) of elliptical and S0 galaxies. The presence of AGN in these early-type galaxies is being investigated by \cite{2015arXiv150703170P} using the stacking technique and has also been found in other samples (e.g. \citealt{2012ApJ...747...57M}), hence in our study we will focus on the non-elliptical dwarf galaxies, where evidence for the presence of a population of accreting BHs remains scarce. After removing the elliptical and S0 galaxies from the sample, the total number of galaxies for our analysis is of 50285 sources. Of these, 97.9\% are starburst and 2.1\% are spiral (Sa-Sdm). Their rest-frame \textit{K}-band luminosity is plotted versus redshift in Figure~\ref{completeness}. The solid black line represents the \textit{K}-band sensitivity limit, for which we consider here a conservative value of 23 mag (the actual value is of 23.7 $\pm$ 0.1 mag; \citealt{2012A&A...544A.156M}). The stellar mass $M_{*}$ is measured by rescaling the best-fit SED, which peaks at \textit{K}-band and is normalized at one solar mass, for the intrinsic luminosity (\citealt{2010ApJ...709..644I}). The distribution of the stellar masses for the sample of low-mass non-early-type galaxies is plotted in Figure~\ref{morphology}, bottom.

\begin{table*}
\begin{minipage}{\textwidth}
\centering
\caption{X-ray stacking results of low-mass star-forming galaxies}
\label{table1}
\begin{tabular}{lccccccccc}
\hline
\hline 
	            & 			& 				    &  \multicolumn{2}{c}{0.5-2 keV}	&  \multicolumn{2}{c}{2-8 keV} \\\cmidrule(r){4-5} \cmidrule(r){6-7}
Bin 		     &	 Exp. time		&	$N_\mathrm{stacked}$ & Detection 			 &  Net count rate & Detection 			&  Net count rate	& log $L_\mathrm{0.5-2 keV}$ & log $L_\mathrm{2-10 keV}$ & log $L_\mathrm{0.5-8 keV}$  \\
		    &	[s]			&	    				&	[$\sigma$]		&   [cts s$^{-1}$]  &		[$\sigma$]	&  [cts s$^{-1}$]	&  [erg s$^{-1}$]  &  [erg s$^{-1}$]  &  [erg s$^{-1}$] \\
    (1)	    &   	(2)    	 &     (3)   				&   (4)     		   		&    (5)    		     & 		(6)   		&     (7)  			&	(8)		  &     (9)   &   (10) \\    
\hline
0 $< z <$ 0.3	& 7.38e+08	&	 7544 			&		5.1	& 8.78e-07$^{1.05e-06}_{6.92e-07}$ 	&	1.6			&	5.28e-07$^{8.53e-07}_{-1.55e-07}$	& 38.92	& 39.37 & 39.43\\ 
0.3 $< z <$ 0.5	& 1.00e+09	&	10187			&		4.1	&5.87e-07$^{7.38e-07}_{4.28e-07}$	&	1.0			&	$<$5.45e-07					& 39.28	& 39.74 & 39.80\\ 
0.5 $< z <$ 0.7	& 9.65e+08	&	10000 			&		3.6	&5.38e-07$^{6.94e-07}_{3.62e-07}$  &	$<$1			&	$<$4.18e-07					& 39.64	& 40.10 & 40.16\\ 
0.7 $< z <$ 1	& 1.08e+09	&	10910			&		4.6	&6.34e-07$^{7.65e-07}_{4.81e-07}$	&	2.5			&	6.68e-07$^{9.74e-07}_{4.14e-07}$	& 40.02	& 40.48 & 40.54\\ 
1 $< z <$ 1.5	& 2.90e+08	&	2957				&		3.7	&9.84e-07$^{1.24e-06}_{6.98e-07}$	&	1.9			&	9.99e-07$^{1.53e-06}_{5.22e-07}$	& 40.54	& 41.00 & 41.06\\ 
$z >$ 1.5$^{*}$	& 5.22e+07	&	514				&		$<$1	&  $<$ 5.94e-07				&	$<$1			& 	$<$3.01e-06					& $<$40.76 & $<$41.22 & $<$41.27 \\ 
\hline
\end{tabular}
\end{minipage}
\raggedright
\smallskip\newline\small {\bf Column designation:}~(1) complete redshift bin; (2) total exposure time; (3) number of stacked galaxies; (4,5) detection significance above the noise and net count rate for the soft (0.5-2 keV) band; (6,7) detection significance above the noise and net count rate for the hard (2-8 keV) band. The errors are provided as upper and lower values at a 68\% confidence level. In the case of no detection above $\gtrsim$2$\sigma$, upper limits are provided as the net count rate at the 68\% confidence level; (8)-(10) 0.5-2 keV, 2-10 keV, and 0.5-8 keV, respectively, unabsorbed X-ray luminosities computed from column (5) assuming $\Gamma=1.4$ and $N_\mathrm{H}$= 2.6 $\times$ 10$^{20}$ cm$^{-2}$. $^*$ This bin is not complete.
\end{table*}

\begin{figure*}
 \includegraphics[width=\textwidth]{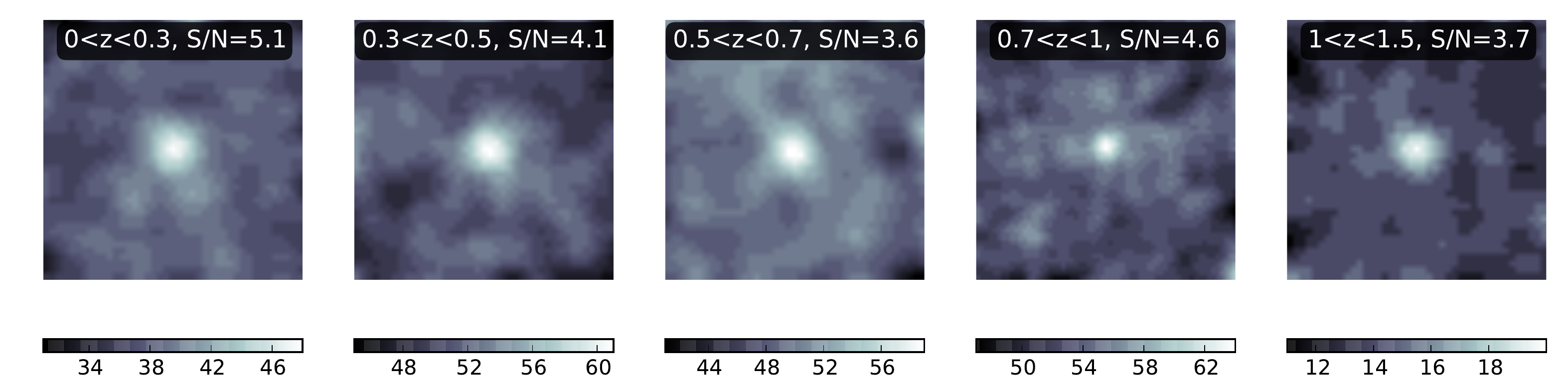}
\protect\caption[detections]{Stacked X-ray detections in the 0.5-2 keV band. Images have been smoothed with a Gaussian of radius = 2. Color scales are in counts.}
\label{detections}
\end{figure*}

\section{X-ray stacking analysis}
\label{analysis}
Only $\sim$1\% of the low-mass galaxies included in our sample are detected above the \textit{Chandra} COSMOS-Legacy flux limit of $F_\mathrm{2-10 keV}$ = 1.7 $\times$ 10$^{-15}$ erg s$^{-1}$ cm$^{-2}$ at 50\% completeness (\citealt{2016arXiv160100941C}; \citealt{2015arXiv151201105M}; M. Mezcua et al. in preparation). For those low-mass non-early-type galaxies with no individual \textit{Chandra} detections, we use the \textit{Chandra} stacking analysis tool \textsc{CSTACK}\footnote{\url{http://lambic.astrosen.unam.mx/cstack_v4.2/}} v4.2 with the aim of unveiling a population of hidden accreting BHs in dwarf starburst and spiral galaxies. 

\textsc{CSTACK} stacks the \textit{Chandra} field of each undetected galaxy, removing any X-ray detected targets and any objects affected by nearby resolved X-ray sources, and returns the exposure-weighted mean X-ray count rates of the stacked population of galaxies in the soft (0.5-2 keV) and hard (2-8 keV) bands by default. \textsc{CSTACK} makes use of the 117 observations (OBSIDs) from \textit{Chandra} COSMOS-Legacy (total of 4.8 Ms of exposure; \citealt{2016arXiv160100941C}) with exposure maps. The \textit{Chandra} COSMOS-Legacy observing strategy is of an highly overlapping (half a field shift) mosaic, where each position is observed with up to 6 \textit{Chandra} observations, therefore at different off-axis angles.
In order to optimize the signal-to-noise (S/N) ratio of the stacked signal, by default \textsc{CSTACK} uses the 90\% energy encircled radius (minimum of 1 arcsec, maximum 7 arcsec) for each of the overlapped observations for an object to calculate the background-subtracted count rate. The background is taken from the 30 $\times$ 30 arcsec$^{2}$ region centered at the object and excluding the 7 arcsec radius circle around the object by default. The source count rate is corrected for the fraction of the point spread function that falls into the source count extraction radius. The exposure-weighted average of these background-subtracted count rates over the input objects over the \textit{Chandra} OBSIDs gives the final stacked count rate. In order to include only datasets that have reasonably small point spread functions, \textsc{CSTACK} includes observations in which the object's off-axis angles are below 8 arcmin by default. While \textsc{CSTACK} allows various options and these defaults can be modified, we use the defaults provided by \textsc{CSTACK} v4.2 in this work.\footnote{See the explanation manual linked to the main entry form of \textsc{CSTACK} for full options. Visitors may login as username=guest, password=guest.} The significance of each stacked detection is determined from the photon counting statistics, while the uncertainties associated to the stacked count rates are evaluated by a bootstrap re-sampling analysis that provides the distribution of the stacked count rates for 500 resampled catalogs, each of which consists of the same number of objects as the input catalog and selected at random from the input catalog allowing duplicates. We note that if the number of photons/object is $\lesssim1$ in the stacking analysis (which is our case), the bootstrap errors become approximately equal to the photon counting statistics.

 \begin{figure}
 \includegraphics[width=0.48\textwidth]{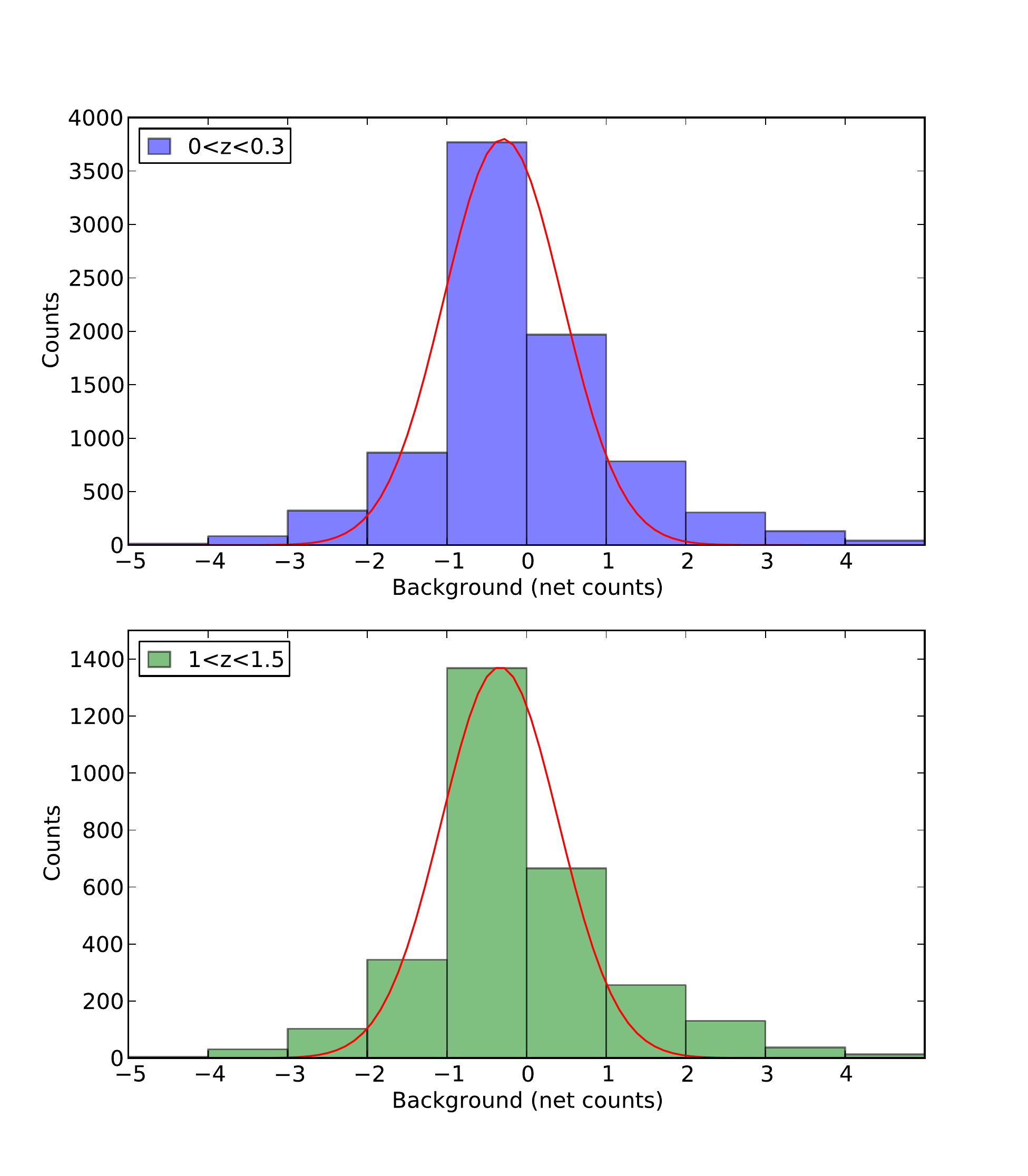}
\protect\caption[background]{Background net count distribution for randomly distributed source positions in the $0 < z < 0.3$ (top) and $1 < z < 1.5$ (bottom) redshift bins. The red solid line shows the fit of a Gaussian distribution with offset = -0.295 and $\sigma$ =  1.052 for the $0 < z < 0.3$ bin and offset = -0.334 and $\sigma$ =  1.037 for the $1 < z < 1.5$ bin.}
\label{background}
\end{figure}

In order to study any redshift evolution of a faint, X-ray undetected, population of low-mass BHs in non-early-type galaxies, the stacking analysis is performed in five complete $L_\mathrm{K}-z$ bins selected as: (1) including galaxies with \textit{K}-band luminosities above the \textit{K}-band sensitivity limit at that redshift and below a maximum \textit{K}-band luminosity $L_\mathrm{K max}$ = 10$^{30}$ erg s$^{-1}$ Hz$^{-1}$ (or $L_\mathrm{K max}$ = 3.7 $\times$ 10$^{10}$ L$_{\odot}$), which is the upper limit where the sample becomes incomplete at low z; (2) having a stacked detection above 3$\sigma$ in at least one X-ray band. The complete redshift bins include sources up to $z$=1.5 and are shown in Figure~\ref{completeness}. To extend the study to even higher redshift, we also stack all those low-mass starburst and late-type galaxies with $L_\mathrm{K}< L_\mathrm{K max}$ and $z>$1.5 despite the lack of completeness in this bin. The number of stacked sources, total exposure time in each redshift bin, significance of the stacked detection, average net count rate and its error computed by the bootstrapping analysis for each redshift bin and X-ray band are provided in Table~\ref{table1}. We consider a detection is significant if the average stacked count rate is above a 3$\sigma$ level. This is the case for the five complete redshift bins in the soft band (see Figure~\ref{detections}), while in the hard band none of the bins is detected above a 2.5$\sigma$ level. 

We thus consider the soft band detections to compute X-ray luminosities. For each of the five bins, the stacked count rates in the 0.5-2 keV band provided by \textsc{CSTACK} are converted to X-ray luminosities in the 0.5-2 keV (soft), 2-10 keV (hard) and 0.5-8 keV (full) bands assuming a power-law photon index $\Gamma=1.4$ and a Galactic column density $N_\mathrm{H}$= 2.6 $\times$ 10$^{20}$ cm$^{-2}$ (\citealt{2005A&A...440..775K}) as used in \textit{Chandra} COSMOS-Legacy (\citealt{2016arXiv160100941C}). For these parameters, the conversion factor from soft count rate to fluxes in the 0.5-2 keV, 2-10 keV and 0.5-8 keV bands are 6.56 $\times\ $10$^{-12}$ erg cm$^{-2}$ s$^{-1}$ cts$^{-1}$, 2.16 $\times\ $10$^{-11}$ erg cm$^{-2}$ s$^{-1}$ cts$^{-1}$ and 1.59 $\times\ $10$^{-11}$ erg cm$^{-2}$ s$^{-1}$ cts$^{-1}$, respectively. The conversion factors are sensitive to the assumed power-law photon index: for $\Gamma$=1 (=2), there is a change of $\sim$4\% ($\sim$8\%) in the soft band, of $\sim$40\% ($\sim$40\%) in the hard band, and of $\sim$4\% ($\sim$8\%) in the full band. For each redshift bin, the X-ray luminosities are brought to rest frame using a mean K-correction factor $(1+z)^{\Gamma -2}$, where we assume the same $\Gamma$ as specified above and $z$ is taken as the mean redshift of each bin. We note that this yields the same results (within the errors) as computing an individual K-correction factor for each source first and averaging it later.

We evaluate any possible background contribution and check the significance of our stacked detections by randomizing the source positions and re-running the stacking. Figure~\ref{background} shows the results for the lowest and highest complete redshift bins. The background net counts ($net$) are computed as $net = cts_{src} - ((cts_{bkg}/(px_{bkg} \times exp_{bkg})) \times (px_{src} \times exp_{src}))$, where $cts_{src}$ are the source counts, $cts_{bkg}$ the background counts, $px_{bkg}$ the background area, $px_{src}$ the source area, $exp_{bkg}$ the background exposure, and $exp_{src}$ the source exposure. They follow a Gaussian distribution centered on zero, as expected from background fluctuations, with offset = -0.295 and $\sigma$ =  1.052 for the $0 < z < 0.3$ bin and offset = -0.334 and $\sigma$ =  1.037 for the $1 < z < 1.5$ bin. This slight asymmetry is not significant as the effects of the skewness are much smaller than other statistical and/or systematic errors like the bootstrap errors (Table~\ref{table1}). Therefore there is no significant background contributing to the stacked X-ray detections.

\section{Results and Discussion}
\label{results}
The stacking analysis finds a significant X-ray detection ($>3\sigma$) in the soft band for the five complete redshift bins, i.e., up to $z$ = 1.5. For each of these bins, and for the X-ray undetected high-$z$ bin ($z >$ 1.5), we average the main galaxy properties (redshift, stellar mass, SFR, galaxy stellar age, $L_\mathrm{K}$; see Table~\ref{table2}) derived from the SED spectral fitting (\citealt{2009ApJ...690.1236I}). 
The X-ray luminosities of the stacked redshift bins range $L_\mathrm{X} = 10^{39}-10^{40}$ erg s$^{-1}$ in the soft band and $L_\mathrm{X} = 10^{39}-10^{41}$ erg s$^{-1}$ in the hard and full bands (Table~\ref{table1}) and are a mixed contribution of the integrated output of X-ray binary (XRB) populations, hot interstellar medium (ISM) gas and nuclear (AGN) emission: $L_\mathrm{X} =  L_\mathrm{XRB} + L_\mathrm{X,hot} + L_\mathrm{AGN}$. The contribution from XRBs and hot ISM gas should thus be removed from the stacked X-ray emission in order to investigate the presence of AGN in the stacked star-forming low-mass galaxies. A description of how these contributions are derived and removed is provided in the next sections.

\subsection{Contribution from X-ray binaries}
\label{XRB}

\begin{figure}
 \includegraphics[width=0.48\textwidth]{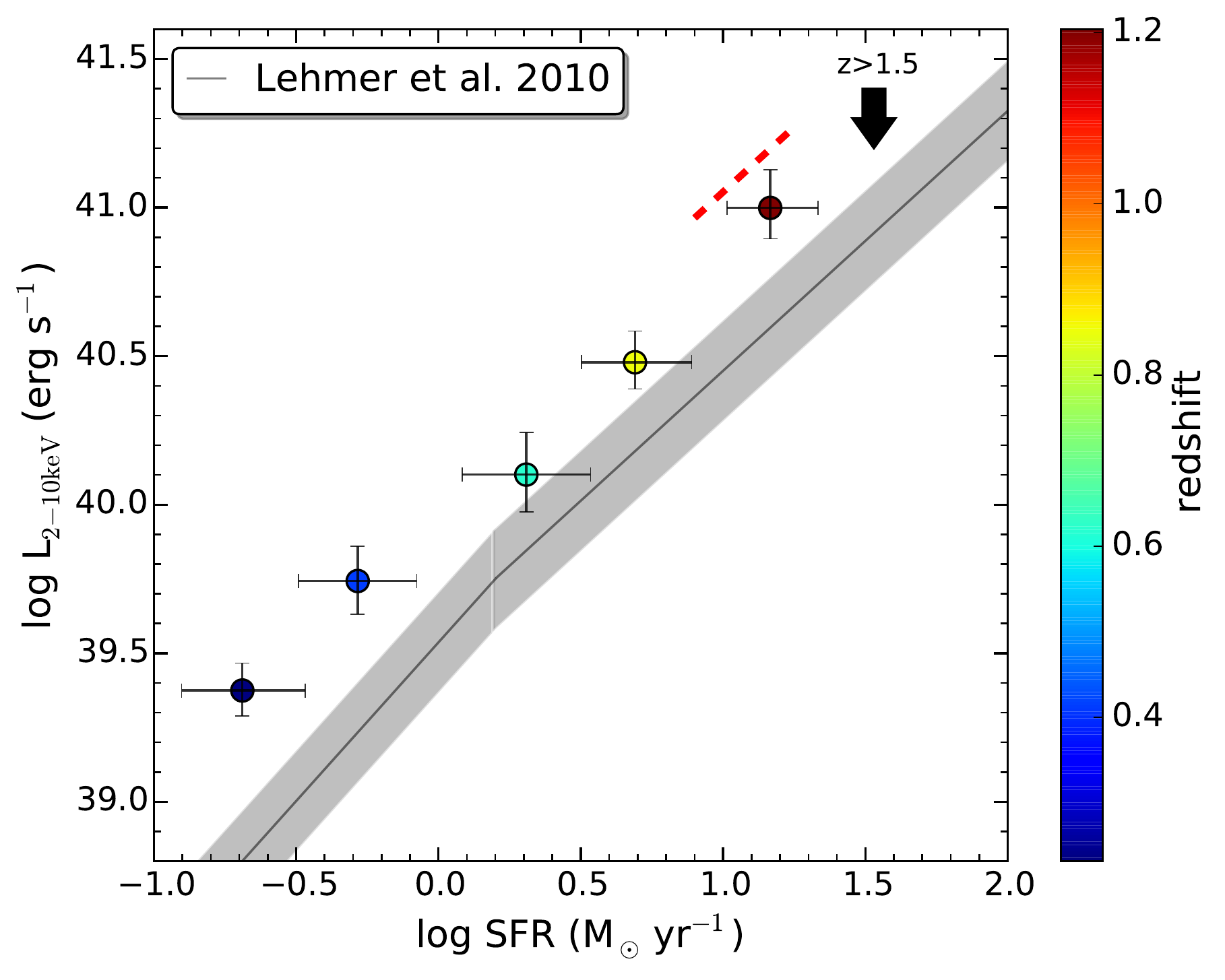}
\protect\caption[completeness]{Luminosity (2-10 keV) versus average SFR of each stacked redshift bin for the sample of starburst and late-type dwarf galaxies. The gray line shows the $L_\mathrm{XRB}$-$M_{*}$-SFR correlation for XRBs from \cite{2010ApJ...724..559L} with a scatter of 0.34 dex. The 1$\sigma$ error bars account for the stacking uncertainties and the statistical errors on the SFRs and M$_{*}$. The red dashed line indicates where the expected contribution from XRBs would lie if the metallicity were a factor three lower than the solar one at z$>$ 1 as predicted by \cite{2013ApJ...776L..31F}.}
\label{SFR}
\end{figure}

To estimate the total contribution from high-mass X-ray binaries (HMXBs) and low-mass X-ray binaries (LMXBs) to the stacked X-ray emission, we use the combined correlation from \cite{2010ApJ...724..559L} for luminous, star-forming galaxies with specific SFRs (sSFR) in the range $\sim$ 2 $\times$ 10$^{8}$ yr$^{-1}$ - 1 $\times$ 10$^{9}$ yr$^{-1}$, which are of the same order as that of our sample of dwarf starburst galaxies (sSFR $\sim$ 10$^{9}$ yr$^{-1}$):
\begin{equation}
\small
L_\mathrm{2-10keV}^\mathrm{XRB} = (9.05 \pm 0.37) \times 10^{28} M_{*} + (1.62 \pm 0.22) \times 10^{39}\ SFR 
\end{equation}
in erg s$^{-1}$ and scatter of 0.34 dex, where the LMXB contribution is proportional to the stellar mass and the HMXB to the SFR. In order to derive and subtract the contribution from XRBs to the stacked soft X-ray emission, the $L_\mathrm{2-10keV}^\mathrm{XRB}$ of each $z$ bin is converted to count rate in the 0.5-2 keV band assuming a photon index $\Gamma$ = 1.4 (which is a good model for the XRB emission) and $N_\mathrm{H}$= 2.6 $\times$ 10$^{20}$ cm$^{-2}$ (see Section~\ref{analysis}) and applying the corresponding K-correction factor. We note that assuming $\Gamma$ = 1.4 we are also taking into account the soft component of XRBs, which is typically hard and approximated with a thermal model with high temperatures ($>$5 keV) corresponding to a power-law model with slope$\sim$1.4 (\citealt{1992ApJS...80..645K}; similar slope derived for the average soft emission of XRBs from table 2 in \citealt{2012MNRAS.426.1870M}). No significant changes are obtained (the differences remain within the errors) by assuming a photon index in the range $\Gamma$=1-2. Using Lehmer's correlation, we find that the contribution from XRBs (including both HMXBs and LMXBs) to the stacked 0.5-2 keV X-ray emission ranges from 9\% for the $1 < z < 1.5$ bin to 15\% for the $0.7 < z < 1$ bin and thus that the stacked X-ray emission is higher than the one expected from XRBs for the five complete redshift bins. This X-ray excess is of $\sim5\sigma$ for the two $z<0.5$ and the $0.7 < z < 1$ bins, $\sim4\sigma$ for the $0.5 < z < 0.7$ bin, and $\sim7\sigma$ for the $1 < z < 1.5$ bin. In the hard band, the contribution from XRBs to the stacked 2-10 keV X-ray luminosity (Table~\ref{table1}) ranges from 15\% for the $0 < z < 0.3$ bin to 27\% for the $0.7 < z < 1$ bin. The $L_\mathrm{X}$ (2-10 keV) obtained from the stacking analysis for each redshift bin is plotted versus SFR in Figure~\ref{SFR}. We note that the stacked $L_\mathrm{X}$ is above the correlation and thus higher than the one expected from XRBs for the five complete redshift bins. This X-ray excess is more significant ($4\sigma$) for $z<0.5$, but it diminishes as we move to higher redshifts and the SFR increases so that for the $0.5 < z < 0.7$, $0.7 < z < 1$ and $1 < z < 1.5$ bins the $L_\mathrm{X}$ is 2-3$\sigma$ above the scatter of the $L_\mathrm{XRB}$-$M_{*}$-SFR correlation. For the highest, incomplete, $z>$1.5 bin, which has the highest value of averaged SFR, the upper limit on the stacked $L_\mathrm{X}$ is $\sim2\sigma$ above that expected from XRBs. However, given that $L_\mathrm{X}$ is an upper limit, all the X-ray emission of this high redshift bin is consistent with that coming from XRBs.

It should be noted that in the low SFR regime, the relation between SFR and the X-ray luminosity from XRBs becomes non linear: $L_\mathrm{XRB} \propto SFR^{1/(\alpha -1)}$; the relation is linear only for high SFRs (\citealt{2004MNRAS.351.1365G}). This could affect the two lowest $z$ bins, for which the SFR is low, decreasing the contribution from XRBs. We test this by computing the contribution from XRBs using $L_{XRB} \propto SFR^{1.6}$ in Lehmer's equation, where we have considered $\alpha$ = 1.6 (the slope of the universal luminosity function of HMXBs; \citealt{2004MNRAS.351.1365G}). We find that the contribution from XRBs  to the stacked 0.5-2 keV X-ray emission diminishes from 12\% for the $0 < z < 0.3$ bin to 5\% and from 11\% for the $0.3 < z < 0.5$ bin to 8\%. These differences are consistent within the errors. We note though that even if the differences were significant, the effect of the non linearity between $L_\mathrm{XRB}$ and SFR for low SFRs is to reduce the expected contribution from XRBs. This would thus increase the resulting X-ray excess, which would be even more significant for the two lowest $z$ bins.

The use of \cite{2010ApJ...724..559L} correlation to estimate the contribution of XRBs to the X-ray emission is common among studies of dwarf galaxies (e.g. \citealt{2013ApJ...773..150S}; \citealt{2015ApJ...805...12L}); nevertheless, other correlations to estimate the contribution from HMXBs and LMXBs separately exist (e.g. \citealt{2003MNRAS.339..793G}; \citealt{2004MNRAS.347L..57G}; \citealt{2012MNRAS.419.2095M}; \citealt{2013ApJ...776L..31F}). To test whether the presence of an X-ray excess is dependent on the correlation used, we estimate the contribution from HMXBs to the stacked X-ray signal using the correlation from \cite{2012MNRAS.419.2095M} for star-forming galaxies. Half of the star-forming galaxies in the parent sample of \cite{2012MNRAS.419.2095M} have $M_{*}\leq3 \times 10^{9}$ M$_{\odot}$ and the correlation is redshift-invariant up to $z$=1.3 (\citealt{2014MNRAS.437.1698M}). For the LMXBs, we estimate their contribution to the stacked X-ray emission using equation 4 in \cite{2013ApJ...776L..31F} and the parameters given in their table 2. Using \cite{2012MNRAS.419.2095M} and \cite{2013ApJ...776L..31F}, the contribution from XRBs (HMXBs+LMXBs) to the stacked 0.5-2 keV X-ray emission ranges from 14\% for the $1 < z < 1.5$ bin to 29\% for the $0.7 < z < 1$ bin and is consistent within the uncertainties with that obtained using the combined \cite{2010ApJ...724..559L} correlation. If instead of \cite{2012MNRAS.419.2095M} to compute the contribution from HMXBs we use equation 3 in \cite{2013ApJ...776L..31F} with metallicity $Z$=0.025 (i.e. the closest value that matches the solar metallicity assumed in the galaxy SED fitting), the results are also consistent within the uncertainties with those found using the \cite{2010ApJ...724..559L} correlation. We thus conclude that, independently of the correlation used, there is an X-ray excess that is more significant at redshift $z < 0.5$ and that in the hard band diminishes as redshift and SFR increase.  

To confirm that the X-ray excess in the two $z<0.5$ bins is not caused by the presence of high star-forming galaxies, we remove those sources with $SFR>3$ M$_{\odot}$ yr$^{-1}$ from the bins and perform the stacking again. By doing this the number of stacked sources reduces to 8043 in the $0<z<0.3$ bin and to 10074 in the $0.3<z<0.5$. We find that the stacked 0.5-2 keV X-ray signal of the $z<0.5$ bins is still 5-6$\sigma$ above the one expected from XRBs, thus confirming that the detections are not dominated by the brightest star-forming galaxies. 

Finally, we note that the lower metallicities expected at higher $z$ for highly star-forming galaxies (e.g. \citealt{2010A&A...521L..53L}; \citealt{2013ApJ...763....9Y}; \citealt{2014ApJ...792...75Z}) suggests a higher contribution of HMXB emission to the integrated $L_\mathrm{X}$ than for the solar metallicity here assumed (\citealt{2013ApJ...776L..31F}), weakening the evidence of AGN emission in the higher $z$ galaxies in our sample. Since we do not have information on the metallicity, to test whether at higher z and for higher SFRs the contribution from HMXBs would be higher than the one reported, we create two subsets of high z ($z>0.7$) galaxies: one with $SFR>6$ M$_{\odot}$ yr$^{-1}$ (6948 sources) and another with $SFR<6$ M$_{\odot}$ yr$^{-1}$ (6919 sources), and perform the stacking and calculate the HMXB contribution again for each subset\footnote{We note that here we choose a division in 6 M$_{\odot}$ yr$^{-1}$ instead of the 3 M$_{\odot}$ yr$^{-1}$ taken for the $z<0.5$ bins because at high redshifts most galaxies are highly star-forming and taking a limit of 3 M$_{\odot}$ yr$^{-1}$ would yield a very reduced number of sources in the $SFR<3$ M$_{\odot}$ yr$^{-1}$ subset compared to the $SFR>3$ M$_{\odot}$ yr$^{-1}$ subset.}. For the $SFR<6$ M$_{\odot}$ yr$^{-1}$ subset, we find that the stacked 0.5-2 keV X-ray luminosity is $L_\mathrm{0.5-2 keV}$ = 39.42 erg s$^{-1}$ while the luminosity expected from XRBs in the 0.5-2 keV band is of $L_\mathrm{XRB}$ = 38.90 erg s$^{-1}$, and that the X-ray excess over $L_\mathrm{XRB}$ is of 4$\sigma$ and thus consistent with the $\sim5\sigma$ previously reported. For the $SFR>6$ M$_{\odot}$ yr$^{-1}$, the stacked 0.5-2 keV X-ray luminosity is $L_\mathrm{0.5-2 keV}$ = 40.51 erg s$^{-1}$, the luminosity expected from XRBs in the 0.5-2 keV band is $L_\mathrm{XRB}$ = 39.52 erg s$^{-1}$, and the X-ray excess is more significant (9$\sigma$). However, this does not mean that the evidence of AGN emission is strengthened at high $z$ and high SFR, as the metallicity in this case is expected to be lower: if the metallicity of the high SFR bin where e.g. a factor of three lower (Z = 0.008), then for the same SFR the contribution from XRBs would increase by a factor 4 (see figure 2 in \citealt{2013ApJ...776L..31F}) and the X-ray excess would disappear. To exemplify this, we show with a red dashed line in Figure~\ref{SFR} where the expected contribution from XRBs would lie for the $z>1$ bin if the metallicity were a factor three lower than the solar one. We thus cannot exclude that the X-ray excess at high $z$ may be due to an increase of the XRB contribution at high SFR and low metallicity.

\subsection{Hot gas contribution}
\label{hot}
In addition to XRBs, the X-ray emitting hot ISM can also contribute to the total emission in star-forming galaxies (e.g. \citealt{2009MNRAS.394.1741O}; \citealt{2013MNRAS.428.2085L}). To remove its contribution from the signal of each stacked bin, we use the correlation between diffuse gas X-ray luminosity ($L_\mathrm{X,hot}$) and SFR from \cite{2012MNRAS.426.1870M}:
\begin{equation}
L_\mathrm{0.5-2keV}^\mathrm{hot} = (8.3 \pm 0.1) \times 10^{38}\ SFR\ \mathrm{(M}_{\odot} \mathrm{yr}^{-1}\mathrm{)} 
\end{equation}
with a scatter of 0.34 dex, and assume a power-law index of $\Gamma=3$ (best representation of a thermal model with temperature $\sim$0.7-1 keV) to convert the $L_\mathrm{0.5-2keV}^\mathrm{hot}$ to soft-band count rate. The corresponding K-correction factor is also applied. We find that the hot ISM contribution to the soft (0.5-2 keV) band stacked X-ray emission ranges from 17\% for the $0 < z < 0.3$ bin to 37\% for the $0.5 < z < 0.7$ bin. After removing this hot gas contribution from the stacked 0.5-2 keV X-ray signal, we find that there is still an X-ray excess above the expected emission from XRBs in the 0.5-2 keV band. This excess has a significance of 4$\sigma$ for the two $z < 0.5$ and the $1 < z < 1.5$ bins and of 2-3$\sigma$ for the $0.5 < z < 0.7$ bin and $0.7 < z < 1$ bins.

\begin{table*}
\begin{minipage}{\textwidth}
\centering
\caption{Host galaxy properties}
\label{table2}
\begin{tabular}{lccccccccc}
\hline
\hline 
Bin		  &	$\bar{z}$ & log $\bar{M_\mathrm{*}}$	&  log $\bar{SFR}$ 				& $\bar{Age}$	& log $\bar{L}_\mathrm{K}$ & log $\bar{L}_\mathrm{hot}$ & log $\bar{L}_\mathrm{XRB}$ & log $\bar{L}_\mathrm{AGN}$ (0.5-2 keV) & log $\bar{L}_\mathrm{AGN}$ (2-10 keV)  \\
		    &		     &	    [$M_\mathrm{\odot}$]	&   [$M_\mathrm{\odot}$ yr$^{-1}$]   &  [Gyr]	&     [erg s$^{-1}$]		&     	 [erg s$^{-1}$]			&     	 [erg s$^{-1}$]	&  [erg s$^{-1}$] &  [erg s$^{-1}$] \\
    (1)	    &   (2)       &     (3)   				&   			(4)     		   	&  	(5)	&	  (6)    		      	 & 		(7)   	               		&    (8)   & (9) & (10)   \\    
\hline
0 $< z <$ 0.3	&	0.23 &		8.38			& 		-0.69				 	&	1.91	&	42.64			&		38.23			& 38.55			& 38.77      & 39.23	 \\ 
0.3 $< z <$ 0.5	&	0.41 &		8.83			& 		-0.28				 	&	1.50 &	43.08			&		38.64			& 38.96			& 39.13      & 39.59	\\ 
0.5 $< z <$ 0.7	&	0.62 &		8.99			& 		0.31				 	&	0.68 &	43.40			&		39.23			& 39.53			& 39.34      & 39.80	\\ 
0.7 $< z <$ 1	&	0.85 &		9.17			& 		0.69				 	&	0.38 &	43.65			&		39.61			& 39.91			& 39.72      & 40.18	\\ 
1 $< z <$ 1.5	&	1.20 &		9.30			& 		1.17				 	&	0.20 &	43.91			&		40.09			& 40.38			& 40.28      & 40.74	\\ 
$z >$ 1.5$^{*}$	&	1.93 &		9.27			& 		1.54				 	&	0.09 &	43.99			&		40.46			& 40.75			& $<$40.26& $<$40.72 \\ 
\hline
\end{tabular}
\end{minipage}
\raggedright
\smallskip\newline\small {\bf Column designation:}~(1) complete redshift bin; mean (2) redshift, (3) stellar mass, (4) star formation rate, (5) galaxy stellar age, (6) \textit{K}-band luminosity, (7) 0.5-2 keV X-ray luminosity expected from hot ISM gas estimated using the correlation from \cite{2012MNRAS.426.1870M}, (8) 2-10 keV X-ray luminosity expected from XRBs estimated using the correlation from \cite{2010ApJ...724..559L}, and (9,10) nuclear X-ray luminosity in the 0.5-2 keV and 2-10 keV, respectively, of the stacked galaxies in each bin after removing the contribution from XRBs and hot ISM gas. $^*$ This bin is not complete.
\end{table*}

\subsection{ULX contribution}
A number of ULXs is also expected to contribute to the stacked X-ray emission in low-mass star-forming galaxies (e.g. \citealt{2008ApJ...684..282S}; \citealt{2011MNRAS.416.1844W}). However, ULXs in star-forming galaxies are mostly the high luminosity end of the HMXB (and to lesser extent LMXB) distribution (e.g. \citealt{2004ApJS..154..519S}; \citealt{2011MNRAS.416.1844W}; see review by \citealt{2011NewAR..55..166F}) and thus their contribution has already been implicitly taken into account (and removed from the stacked X-ray emission) when computing the XRB contribution. We note though that there exists a small fraction of ULXs, those with X-ray luminosities above $5 \times 10^{40}$ erg s$^{-1}$, that can difficulty be explained by stellar-mass objects and constitute the best candidates to IMBHs (e.g. \citealt{2011MNRAS.416.1844W}; \citealt{2012MNRAS.423.1154S}). The two strongest cases so far found have been suggested to be the nucleus of stripped dwarf galaxies (e.g. \citealt{2012ApJ...747L..13F}; \citealt{2013ApJ...768L..22S}; \citealt{2015MNRAS.448.1893M}), and thus if present in the galaxies here studied they would contribute to the nuclear X-ray emission.

\subsection{Nuclear X-ray emission}
\label{nuclear}
The stacked X-ray emission of the sample of low-mass non-early-type galaxies has a significant contribution from XRBs and diffuse hot gas emission. What is left after removing these XRB and hot ISM contributions is most likely nuclear X-ray emission, indicating the possible presence of accreting BHs. To derive the AGN emission, we subtract the XRB and hot ISM contributions obtained in Sections~\ref{XRB} and \ref{hot} from the stacked soft count rates and convert them to 0.5-2 keV and 2-10 keV luminosities (properly K-corrected) assuming $\Gamma$ = 1.4 --which is a good assumption for AGN emission and it is also the slope of the cosmic X-ray background (e.g. \citealt{2006ApJ...645...95H}) and therefore a good representation of a mixed distribution of obscured and unobscured sources-- and $N_\mathrm{H}$= 2.6 $\times$ 10$^{20}$ cm$^{-2}$ (see Section~\ref{analysis}).
We find that the AGN X-ray luminosities are still in the range $L_\mathrm{AGN} = 10^{39}-10^{40}$ erg s$^{-1}$ in the soft band and $L_\mathrm{AGN} = 10^{39}-10^{41}$ erg s$^{-1}$ in the hard band (see Table~\ref{table2}), and thus above the typical X-ray luminosity of stellar-mass BHs and globular clusters (e.g. \citealt{2007ApJ...661..875K}; \citealt{2010MNRAS.406.1049C}; \citealt{2012ApJ...750L..27S}) but one to two orders of magnitude lower than the typical X-ray luminosity limit considered for AGN ($10^{42}$ erg s$^{-1}$).

The presence of a population of accreting BHs in low-mass starburst and spiral galaxies is more significant (4$\sigma$) for $z<$0.5, where the $L_\mathrm{AGN}$ fraction (defined as the fraction of X-ray excess that contributes to the stacked X-ray luminosity once the $L_\mathrm{XRB}$ and $L_\mathrm{X,hot}$ contributions have been removed; i.e. AGN fraction = $L_\mathrm{AGN}/L_\mathrm{X}*100)$ is of $\sim$70\%, while the three highest $z$ bins have $L_\mathrm{AGN}$ fractions of $\sim$50\%. This does not rule out the existence of high-$z$ dwarf starburst galaxies hosting AGN. At $z > 0.5$, when the SFR is higher, the BHs might be obscured or hidden, as found by \cite{2012ApJ...758..129X} for blue low-mass galaxies and also suggested by \cite{2014ApJ...790...16C} and \cite{2015arXiv150703170P} for early-type galaxies with an excess of X-ray emission with respect to their $L_\mathrm{K}$. We investigate this in the next section by means of the hardness ratio (HR). 

\begin{figure}
 \includegraphics[width=0.48\textwidth]{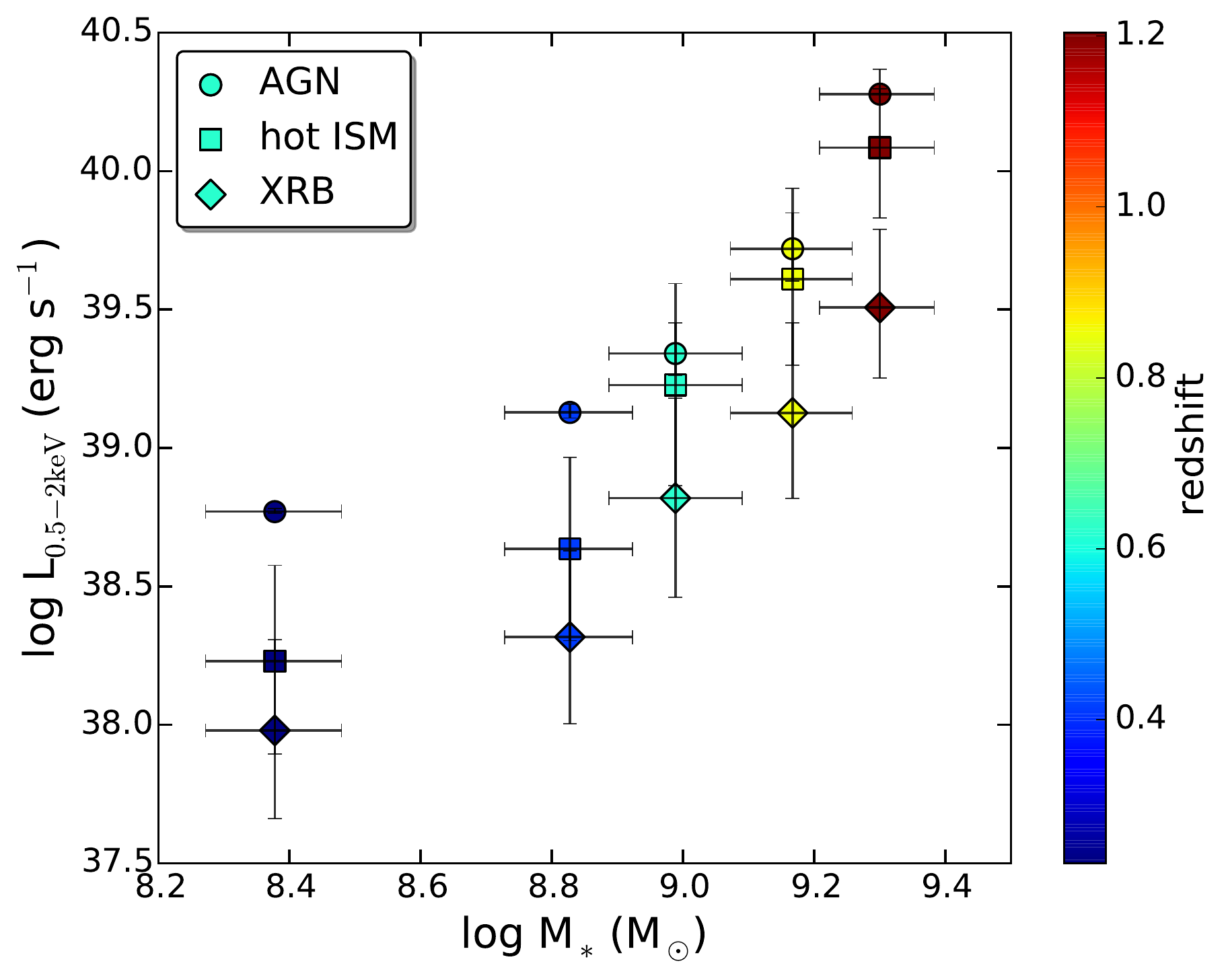}
\protect\caption[completeness]{Nuclear X-ray luminosity (circles), hot ISM luminosity (squares) and XRB luminosity (diamonds) in the 0.5-2 keV band vs. average stellar mass of each stacked redshift bin.}
\label{AGN}
\end{figure}

In Figure~\ref{AGN} we show how $L_\mathrm{AGN}$, $L_\mathrm{XRB}$ and $L_\mathrm{X,hot}$ vary with $M_\mathrm{*}$ and $z$. The lack of sources in the top left (high luminosity and low mass) and bottom right corners (low luminosity and high mass) of this figure is mainly due to sample biases. Dwarf spiral and irregular galaxies with bright detected nuclear emission ($L_\mathrm{X} > 10^{40}$ erg s$^{-1}$), similar to those presented by e.g. \cite{2011Natur.470...66R}, \cite{2013ApJ...773..150S}; \cite{2014ApJ...787L..30R}, \cite{2015ApJ...809L..14B} and \cite{2015ApJ...798...38S}, would be located in the upper left side of the figure (e.g. the region with $L_\mathrm{X} > 10^{40}$ erg s$^{-1}$ for log $M_\mathrm{*} <  9$). These X-ray detections will be presented in a forthcoming work (Mezcua et al. in preparation). 
The lack of low luminosity and high mass sources is due to the mass limit of the COSMOS optical/infrared survey. The increase of nuclear luminosity with stellar mass could also be due to the fact that at high SFRs (and high $z$) it is more difficult to measure the AGN contribution while at low SFR (and low $z$) the AGN contribution is more easily detectable and thus more significant.


\subsection{Hardness ratio}
\label{hardness}
\begin{figure}
 \includegraphics[width=0.48\textwidth]{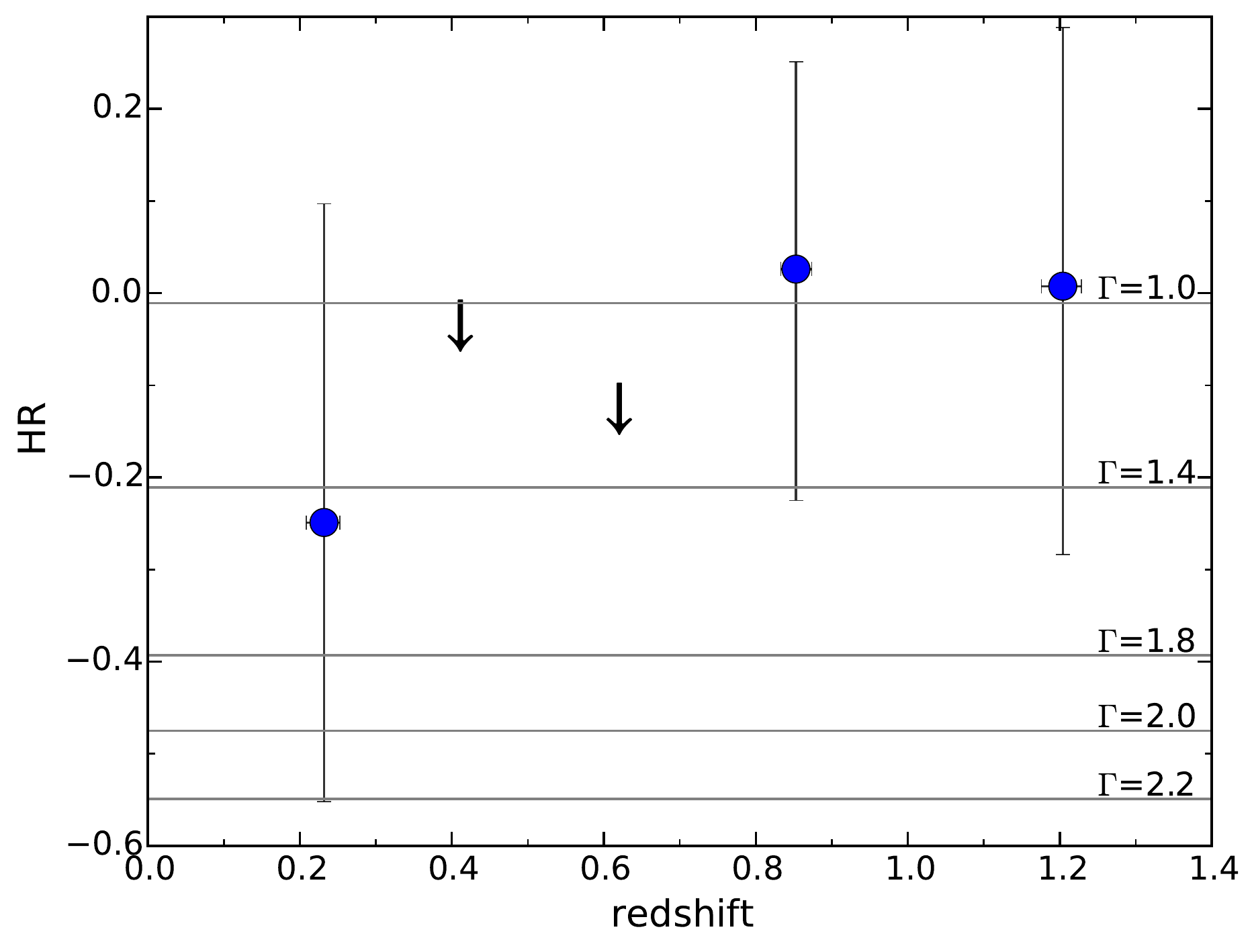}
\protect\caption[hardness]{Hardness ratio vs. redshift for the five complete redshift bins. The lines indicate the HR from a population of XRBs with $\Gamma$ = 1, 1.4, 1.8, 2, 2.2 (from top to bottom) and Galactic $N_\mathrm{H}$.}
\label{hardnessratio}
\end{figure}

The X-ray excess of the five complete $z$ bins with respect to the X-ray emission expected from XRBs and hot ISM suggests the presence of hidden AGN in the population of low-mass starburst and spiral galaxies. To test this, we derive the X-ray HR defined as HR = ($H$-$S$)/($H$+$S$), where $H$ and $S$ are the count rates in the soft (0.5-2 keV) and hard (2-8 keV) \textit{Chandra} bands, respectively. Although the stacking analysis provided significant detections ($>3\sigma$) only in the soft band, for the purpose of this test we consider as a detection those hard-band count rates detected at a $\gtrsim$ 2$\sigma$ level (see Table~\ref{table1} and Figure~\ref{hardnessratio}). These HR are in the observed frame. For a power-law spectral slope they are $z$-invariant (see Figure~\ref{hardnessratio}).

Unobscured AGN showing unabsorbed soft spectra have typically $HR<-0.1$ (e.g. \citealt{2008A&A...490..905H}; \citealt{2012ApJS..201...30C}). This is the case for the lowest redshift bin ($z<0.3$), while the higher redshift bins ($z >0.7 $) have HR $>$ 0, supporting the presence of obscured AGN. To further test the presence of AGN in the stacked galaxies, we model the $HR$ of different populations of XRBs and check whether a hard component is still required to match the HR of the dwarf non-elliptical galaxies. For this, we plot tracks at a constant Galactic column density $N_\mathrm{H} = 2.6 \times 10^{20}$ cm$^{-2}$ and photon indices varying from $\Gamma$ = 1 to $\Gamma$ = 2.2 (see Figure~\ref{hardnessratio}). The $HR$ of the five stacked complete redshift bins are located in a region close to the $\Gamma$=1 model and the $\Gamma$ = 1.4 model, which points to the presence of hard X-ray emission and supports the presence of AGN in the dwarf galaxies here studied. We note that this range of spectral slopes is also consistent with that assumed to convert the stacked count rates to fluxes (Sections~\ref{analysis} and \ref{nuclear}). A slight tendency toward higher HR and harder photon indices (i.e. higher column densities) for higher redshifts is also observed, as was found by \cite{2014ApJ...783...25J} for a $K$-band selected sample of galaxies with $0.5 < z < 2$ in the COSMOS survey, in agreement with the finding of a lower X-ray excess for the higher $z$-bins (Section~\ref{nuclear}).

\subsection{Black hole mass and accretion rate}
\label{BHmass}
The formation of BH seeds via direct collapse is the least stringent and most widely supported scenario (e.g. \citealt{2013ApJ...771..116J}; \citealt{2014GReGr..46.1702N}), which results in massive seeds of $M_\mathrm{BH} \sim10^{4}-10^{5}$ M$_\mathrm{\odot}$. The best IMBH candidates so far detected in the nearby Universe have BH masses in this regime (e.g. \citealt{2012Sci...337..554W}; \citealt{2014Natur.513...74P}; \citealt{2015ApJ...809L..14B}; \citealt{2015MNRAS.448.1893M}), while the detection of lighter seed BHs (PopIII formation scenario) remains elusive (the best candidate so far is CR7; \citealt{2015ApJ...808..139S}). To get an estimate of the mean BH mass in our stacked sample, we assume a bolometric correction of $k$ = $L_\mathrm{bol}/L_\mathrm{X}=5$ (i.e. in between those of AGN and XRBs; \citealt{2015MNRAS.448.1893M}). Using $M_\mathrm{BH} = (5\times L_\mathrm{X}/l_\mathrm{Edd})/1.3\times10^{38}$ M$_{\odot}$, where $l_\mathrm{Edd}=L_\mathrm{bol}/L_\mathrm{Edd}$ is the Eddington ratio and we take $L_\mathrm{X}$ as the $L_\mathrm{AGN}$ in the 0.5-2 keV band, and assuming $l_\mathrm{Edd} \geq 10^{-2}$ as that found in optically-selected samples of low-mass BHs (e.g. \citealt{2004ApJ...610..722G,2007ApJ...670...92G}; \citealt{2012ApJ...755..167D}; \citealt{2013ApJ...775..116R}; \citealt{2014ApJ...782...55Y}; \citealt{2015ApJ...809L..14B}) would yield BH masses $M_\mathrm{BH} \leq  1 \times 10^{5}$ M$_{\odot}$ for the five complete redshift bins, consistent with the IMBH regime. Assuming these BH masses and the median stellar mass as a proxy for the bulge mass, the sources would be 100-400 times off the $M_\mathrm{BH}$-$M_\mathrm{bulge}$ correlation (e.g. \citealt{2013ApJ...764..184M}; see also e.g. \citealt{2012NatCo...3E1304G}; \citealt{2015ApJ...809L..14B}) but consistent with the steeper relation at low masses of \cite{2013ApJ...768...76S} and \cite{2015ApJ...798...54G}.

Vice versa, assuming that the population of accreting BHs in each complete redshift bin falls on the $M_\mathrm{BH}$-$M_\mathrm{bulge}$ correlation (with $M_\mathrm{bulge}$=$M_\mathrm{*}$), the BH masses would be $M_\mathrm{BH} \leq 7 \times 10^{5} - 6 \times 10^{6}$ M$_{\odot}$, which is closer to the typical mass limit considered for SMBHs (10$^{6}$ M$_{\odot}$; e.g. \citealt{2004ApJ...610..722G}), as has been also found for other accreting BHs in dwarf galaxies (e.g. \citealt{2011Natur.470...66R,2014ApJ...787L..30R}). This does not rule out the presence of IMBHs. The accretion would be highly sub-Eddington, with $l_\mathrm{Edd} \leq 2 \times 10^{-4}$, consistent with the findings for low-luminosity AGN with $M_\mathrm{BH} > 10^{6}$ M$_{\odot}$ and $l_\mathrm{Edd}$ = 10$^{-7}-10^{-2}$ (e.g. \citealt{2008ARA&A..46..475H}; \citealt{2011A&A...527A..23M}; \citealt{2014ApJ...787...62M}). The upper limit reflects the fact that most of the low-mass galaxies here studied are classified as starburst while the BH mass scaling relations are typically calibrated for early- and late-type galaxies (e.g. \citealt{2013ApJ...764..184M}). Indeed, assuming $M_\mathrm{bulge}$=$M_\mathrm{*}$ to estimate BH masses may yield an overestimate of more than one order of magnitude (e.g. \citealt{2015ApJ...813...82R}). To reduce this effect, we consider instead the correlation between BH mass and stellar mass found in local AGN by \cite{2015ApJ...813...82R} and which includes a sample of dwarf galaxies:
log ($M_\mathrm{BH}/$M$_{\odot}$) = 7.45 $\pm$ 0.08 + (1.05 $\pm$ 0.11) log ($M_\mathrm{*}/10^{11}$ M$_{\odot}$), with a scatter of 0.55 dex. Using this correlation, the BH masses range $M_\mathrm{BH} = 1 - 9 \times 10^{5}$ M$_{\odot}$ and are again consistent with IMBHs. 
The Eddington ratios in this case are within the range $l_\mathrm{Edd} = 9 \times 10^{-4} - 1 \times 10^{-3}$, lower than the values reported for optically selected samples of low-mass BHs, but consistent with those of low-luminosity AGN (e.g. \citealt{2014ApJ...787...62M}). 
 
Last, we note that the BH mass inferred above could be higher due to a fundamental limitation of the stacking analysis, which is that it does not allow us to discriminate between cases where a small fraction of objects emit X-rays strongly and the rest much weaker and where all the objects emit X-rays at the average level. This is, if AGN follow a duty cycle (e.g.  \citealt{2001ApJ...547...27H}; \citealt{2001ApJ...547...12M}) in which a fraction of their time $f$ they are ``on'' (``AGN-on'', with e.g. $l_\mathrm{Edd} > 0.01$), then a fraction $f$ of all low-mass non-elliptical galaxies that meet our sample selection criteria have AGN ``on'' (as $f\sim\frac{\text{\# sample galaxies with AGN-on}}{\text{total \# of sample galaxies}}$; \citealt{2001ApJ...547...27H}; \citealt{2001ApJ...547...12M}) while the rest of them are ``off'' (e.g. $l_\mathrm{Edd} << 0.01$). The stacking analysis results and the subtraction of the star formation components gives an estimate of the AGN luminosity averaged over all non-elliptical galaxies ($\langle L_{\rm X,AGN}\rangle_\mathrm{on+off}$). If this AGN contribution only comes from the fraction $f$ of all sample galaxies, the average X-ray luminosity of ``on'' galaxies becomes  $\langle L_\mathrm{X,AGN} \rangle_\mathrm{on}=\langle L_\mathrm{X,AGN}\rangle_\mathrm{on+off}/f$ so that if e.g. $f=10$\%, then the average luminosity of AGNs would increase by a factor 10 and the BH mass estimate would become 10 times larger.

\section{Conclusions}
\label{conclusions}
We have studied a population of low-mass ($M_{*}\leq3 \times 10^{9}$ M$_{\odot}$) star-forming galaxies drawn from the COSMOS survey with the aim of investigating the presence of accreting BHs and putative seed BHs from which SMBHs grow. By performing stacking analysis using \textit{Chandra} COSMOS-Legacy images at the position of each non-detected galaxy, we find a significant X-ray detection in each of five complete redshift bins spanning from $z=0$ to $z=1.5$. After removing the contribution of HMXBs, LMXBs, and hot ISM gas to the stacked X-ray luminosity, we still find an X-ray excess in all redshift bins. This X-ray excess is more significant ($4\sigma$) for $z<0.5$ and can be accounted for by nuclear accreting BHs. The average nuclear X-ray luminosities in the 2-10 keV band range from 2 $\times 10^{39}$ erg s$^{-1}$ for the $0 < z < 0.3$ bin to $6 \times10^{40}$ erg s$^{-1}$ for the $1 < z < 1.5$ bin. 
At high redshift and for larger stellar masses and SFRs, the putative accreting BHs could be obscured, as supported by the finding of hard HRs but no significant hard-band X-ray emission. If accreting at Eddington ratios exceeding $10^{-2}$, the BHs would have masses in the intermediate regime but 100-400 times higher than those implied by the local $M_\mathrm{BH}$-$M_\mathrm{bulge}$ correlation. The location of the sources would be consistent with the BH mass-stellar mass scaling relations if they had Eddington ratios $\sim10^{-3}$ and BH masses $\sim 10^{5}$ M$_{\odot}$. 

We conclude that a population of low-mass accreting BHs in dwarf star-forming galaxies similar to those seed BHs populating the early Universe exists. Given their faintness (X-ray luminosities of the same order as that of local ULXs), highly sub-Eddington accretion rates, and obscuration as the redshift and SFR increase, the detection of these BHs is challenging even in \textit{Chandra}  deep surveys. The wider-area \textit{Chandra} COSMOS Legacy survey, despite being shallower than the \textit{Chandra} Deep Fields, has allowed the stacking of a larger number of sources providing the significant stacked detections shown in this paper. In a following work, we will discuss the $\sim$200 dwarf star-forming galaxies that are individually detected in the \textit{Chandra} COSMOS Legacy survey.

\section*{Acknowledgments}
We thank the referee Jenny Greene for her valuable comments which have helped improve this manuscript. M.M. acknowledges financial support from NASA Chandra Grant G05-16099X. This work was supported in part by NASA Chandra grant number GO7-8136A (F.C. and S.M.). F.C. is gratefull to Debra Fine for her support to women in science. T.M. acknowledges support from UNAM-DGAPA Grant PAPIIT IN104113 and  CONACyT Grant Cient\'ifica B\'asica \#179662. The development of CSTACK was partially supported by the Chandra Guest Observer Support Grant GO1-12178X.

\bibliographystyle{apj} 
\bibliography{/Users/mmezcua/Documents/referencesALL}

\end{document}